# High-Speed and High-Responsivity Asymmetric Waveguide Photodiode with Low Optical Back-Reflection

**Zhijun Zhang[1,†,a)], Xuejie Gao[1,2,†], Qiunan Li[1,3], and Xiaoyu Mi[1,b)]**

**AFFILIATIONS**

[1]Yongjiang Laboratory, Ningbo, China

[2]Beijing University of Posts and Telecommunications, Beijing Key Laboratory of MEMS Technology and Device Reliability for Industrial Internet, Beijing, China

[3]School of Physics University of Electronic Science and Technology of China Chengdu, China

[†]These authors contributed equally to this work.

[a)]**Author to whom correspondence should be addressed**: zhijun-zhang@ylab.ac.cn

[b)]xiaoyu-mi@ylab.ac.cn

**ABSTRACT**

We numerically demonstrate an asymmetric corner-reflector uni-traveling-carrier waveguide photodiode (UTC-WGPD) that simultaneously suppresses optical back-reflection and breaks the bandwidth-responsivity trade-off. Off-center beam coupling introduces geometric asymmetry to eliminate retroreflection while maintaining efficient light trapping. 3D optoelectronic simulations reveal that at 1550 nm, a 15-μm$^2$ device exhibits a 275-GHz 3-dB bandwidth, 0.68-A/W responsivity (54.5% quantum efficiency), and -29-dB back-reflection, achieving a 150-GHz bandwidth-efficiency product. This design offers a scalable, low-reflection detection scheme for sub-terahertz transceivers and dense photonic integration.

The proliferation of data-intensive services, including cloud computing, 5G/6G communications, and generative artificial intelligence, is driving rapid growth in global data traffic and an escalating demand for transmission capacity. To meet this demand, per-channel symbol rates in optical links are rapidly scaling from 100 Gbaud toward 200 Gbaud and beyond.[1] In such ultra-high-speed links, the photodiode (PD) serves as the critical optical-to-electrical (O/E) conversion component at the receiver. Its radio-frequency (RF) bandwidth fundamentally limits the achievable symbol rate, while its optical responsivity directly dictates the receiver sensitivity.[2-4]

While surface-illuminated photodetectors are widely employed,[5-8] they suffer from an intrinsic bandwidth-responsivity trade-off. Since their optical propagation and carrier transport axes are collinear, increasing the absorption-layer thickness for higher responsivity inevitably elongates the electron transit distance, degrading the RF bandwidth. In contrast, waveguide photodiodes (WGPDs) circumvent this limitation through an orthogonal geometry.[9-17] By decoupling the optical interaction length from the carrier transit distance, WGPDs enable a thin active region for maximum bandwidth while maintaining a long propagation path for high responsivity.

Nevertheless, as target bandwidths scale beyond 100 GHz toward the sub-terahertz regime, the resistance-capacitance (RC) time constant emerges as the dominant limitation.[14,18] Minimizing the junction capacitance necessitates a reduced active area, which inherently truncates the optical interaction length in WGPDs. Consequently, this curtailed absorption path compromises photon capture, inevitably degrading both external quantum efficiency and responsivity.

To mitigate responsivity degradation in compact WGPDs, reflective structures are typically employed to recycle unabsorbed light for a second pass.[19-22] However, conventional approaches entail inherent trade-offs. Distributed Bragg reflectors (DBRs)[20,21] are wavelength-dependent and impose stringent deep-etching tolerances, restricting their broadband applicability. Conversely, symmetric reflectors[19,22] offer broadband operation but directly retroreflect light into the input waveguide, thereby exacerbating the relative intensity noise of external lasers. Although angled or polygonal geometries[23] can suppress back-reflection to -36 dB via spatial asymmetry, they scatter residual light away from the active region. Without a second absorption pass, such configurations fail to enhance device responsivity.

Here, we propose an asymmetric corner-reflector waveguide photodiode (ACR-WGPD). By synergistically integrating off-center waveguide coupling, a corner reflector, and a lateral light-dissipation facet, this architecture recycles unabsorbed light while geometrically disrupting the longitudinal retroreflective path. Our numerical evaluations indicate that the ACR-WGPD effectively overcomes the bandwidth-responsivity limitation inherent to ultracompact footprints. Specifically, the device delivers a 3-dB bandwidth of 275 GHz and a responsivity of 0.68 A/W within a 15-μm² active area. Furthermore, this geometric asymmetry suppresses optical back-reflection to -28 dB, which mitigates device-level optical feedback and significantly relaxes the stringent demands on external optical isolators.

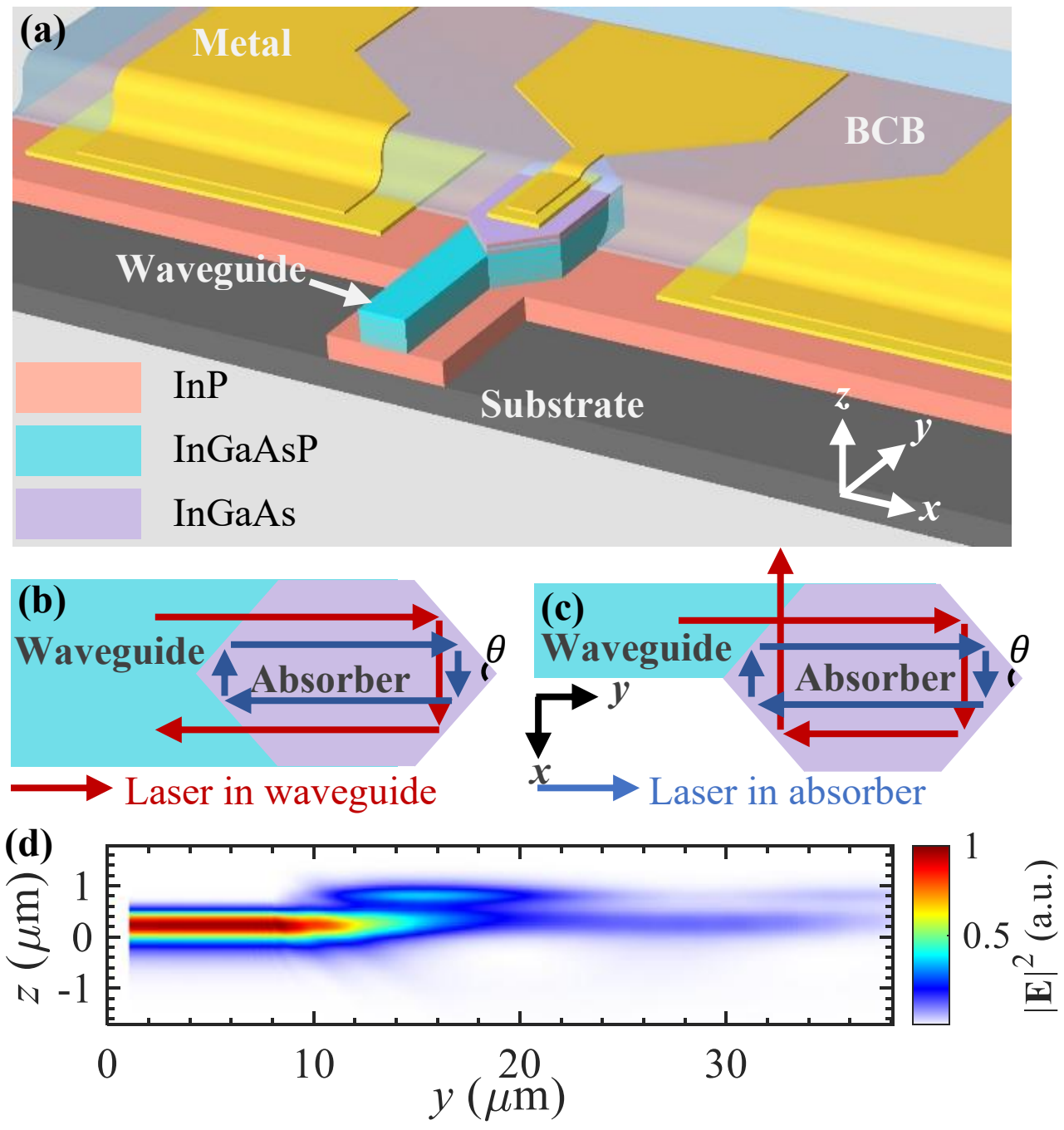

**FIG. 1**. (a) Three-dimensional (3D) schematic of the proposed ACR-WGPD. Top-view schematics illustrating the active regions and optical propagation paths of (b) the conventional CR-WGPD and (c) the proposed ACR-WGPD. (d) Simulated cross-sectional distribution of the optical-field intensity ( $|\mathbf{E}|^2$ ), illustrating evanescent coupling from the passive waveguide into the absorber layer.

In typical evanescently coupled WGPDs as shown in Figs. 1(a) and 1(d), the primary optical mode propagates within the underlying passive waveguide while its evanescent field penetrates the active layer. This orthogonally decouples optical absorption (longitudinal, $y$-axis) from carrier transport (vertical, $z$-axis), enabling a thin active region for minimal transit time without sacrificing the interaction length.

To overcome the scaling limitations of ultracompact designs, the proposed ACR-WGPD synergistically integrates three structural features: off-center waveguide coupling, a corner reflector, and a lateral light-dissipation facet, as shown in Fig. 1(c).

The ACR-WGPD employs an off-center waveguide-coupling scheme. In contrast to the conventional symmetric corner-reflector WGPD (CR-WGPD) as shown in Fig. 1(b), the input waveguide of the ACR-WGPD is laterally offset along the $x$-axis. The incident optical mode then undergoes its first absorption pass as it propagates forward along the $y$-axis.

To recycle unabsorbed light, a corner reflector comprising two facets that meet at an angle $\theta$ is integrated at the rear end of the active region, as shown in Fig. 1(c). The two facets induce successive total internal reflection (TIR), reversing the propagation direction by 180°. Due to the initial waveguide offset, the reflected beam experiences a lateral spatial shift. This folded beam then propagates through the adjacent half of the active layer for a second absorption pass. Consequently, the folded optical path enhances the effective interaction length without increasing the device footprint or junction capacitance.

Symmetric designs retroreflect this folded beam directly into the input waveguide, causing severe optical feedback. In contrast, the ACR-WGPD interrupts the return path using a tilted light-dissipation facet at the front end. The laterally shifted beam impinges on this facet and is redirected along the $x$-direction into the surrounding cladding or substrate. Ultimately, this asymmetric architecture leverages multipass absorption to maintain high responsivity in compact devices, while geometrically eliminating the direct back-reflection path.

Prior to analyzing the optical-field manipulation, we establish the physical constraints governing the device footprint. In ultrahigh-speed photodiodes, the overall 3-dB bandwidth ($f_{3\text{dB}}$) is jointly limited by the carrier transit time and the resistance-capacitance (RC) time constant. To maximize the transit-time-limited bandwidth, a uni-traveling-carrier (UTC) epitaxial structure is adopted [15]. Based on this, an equivalent circuit model[24] incorporating both the carrier-transit-time effect and the parasitic RC network is employed. This model extracts the S-parameters to evaluate the overall frequency response as a function of the device dimensions.

As shown in Fig. 2, with the carrier transit time minimized by the modified UTC structure, the junction capacitance becomes the dominant bandwidth limitation. This capacitance, comprising the absorber ( $C_A$ ) and collector ($C_C$) components, scales linearly with the active area for a fixed epitaxial profile. Consequently, achieving a 3-dB bandwidth ($f_{3dB}$) beyond 250 GHz dictates a stringent reduction in device footprint, as illustrated by the simulated bandwidth-area relationship. This electrical constraint limits the maximum allowable active area to approximately 19 μm$^2$, and a typical active area of 15 μm$^2$ delivers a 3-dB bandwidth of 275 GHz.

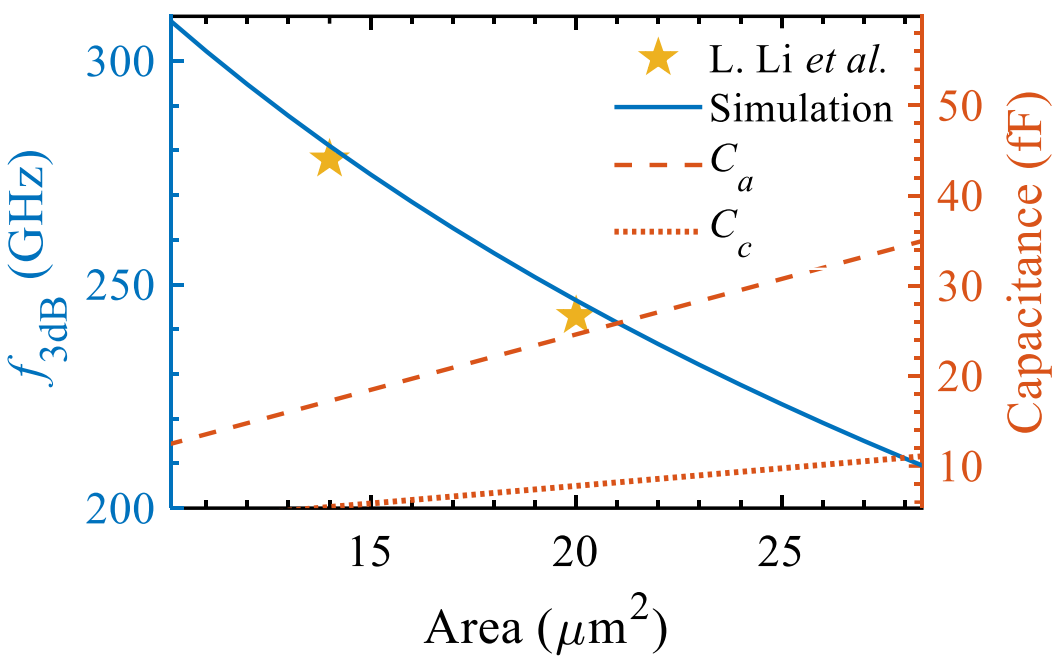

**FIG. 2**. Simulated 3-dB bandwidth ($f_{3dB}$) and extracted junction capacitances as functions of the device active area. $C_a$ and $C_c$ denote the absorber and collector capacitances, respectively. The star symbols represent the experimental data reported in Ref. [15].

As shown in Fig. 2, with the carrier transit time minimized by the modified UTC structure, the junction capacitance becomes the dominant bandwidth limitation. This capacitance, comprising the absorber ($C_A$) and collector ($C_C$) components, scales linearly with the active area for a fixed epitaxial profile. Consequently, achieving a 3-dB bandwidth ($f_{3dB}$) beyond 250 GHz dictates a stringent reduction in device footprint, as illustrated by the simulated bandwidth-area relationship. This electrical constraint limits the maximum allowable active area to approximately 19 μm$^2$, and a typical active area of 15 μm$^2$ delivers a 3-dB bandwidth of 275 GHz.

To quantitatively evaluate the ACR-WGPD architecture, we performed numerical simulations for devices with active areas spanning 13 to 18 μm$^2$. The underlying models employed an InP/InGaAsP material platform at a 1550-nm operating wavelength. The absorption coefficient ($\alpha$) of the InGaAs layer was set to 7000 cm$^{-1}$. Unless otherwise specified, all optical simulations assumed fundamental transverse-electric (TE) mode excitation.

To evaluate the optical performance of the proposed architecture, finite-difference time-domain (FDTD) simulations were performed for three device configurations shown in Fig. 3(a-c): the reference waveguide photodiode (Ref-WGPD), the symmetric CR-WGPD, and the proposed ACR-WGPD.

The primary optical llimitation in ultracompact photodiodes is incomplete single-pass absorption due to the shortened active region. Consequently, the Ref-WGPD suffers from low responsivity, as shown in Fig. 3(d). Furthermore, unabsorbed light propagates longitudinally, reflects off the flat rear facet, and couples directly back into the input waveguide. This reflected field interferes with the forward-propagating mode, generating Fabry-Perot (FP) interference and a longitudinal standing-wave pattern. As a result, the Ref-WGPD exhibits a baseline optical back-reflection between -22 and -17 dB, as shown in Fig. 3(e).

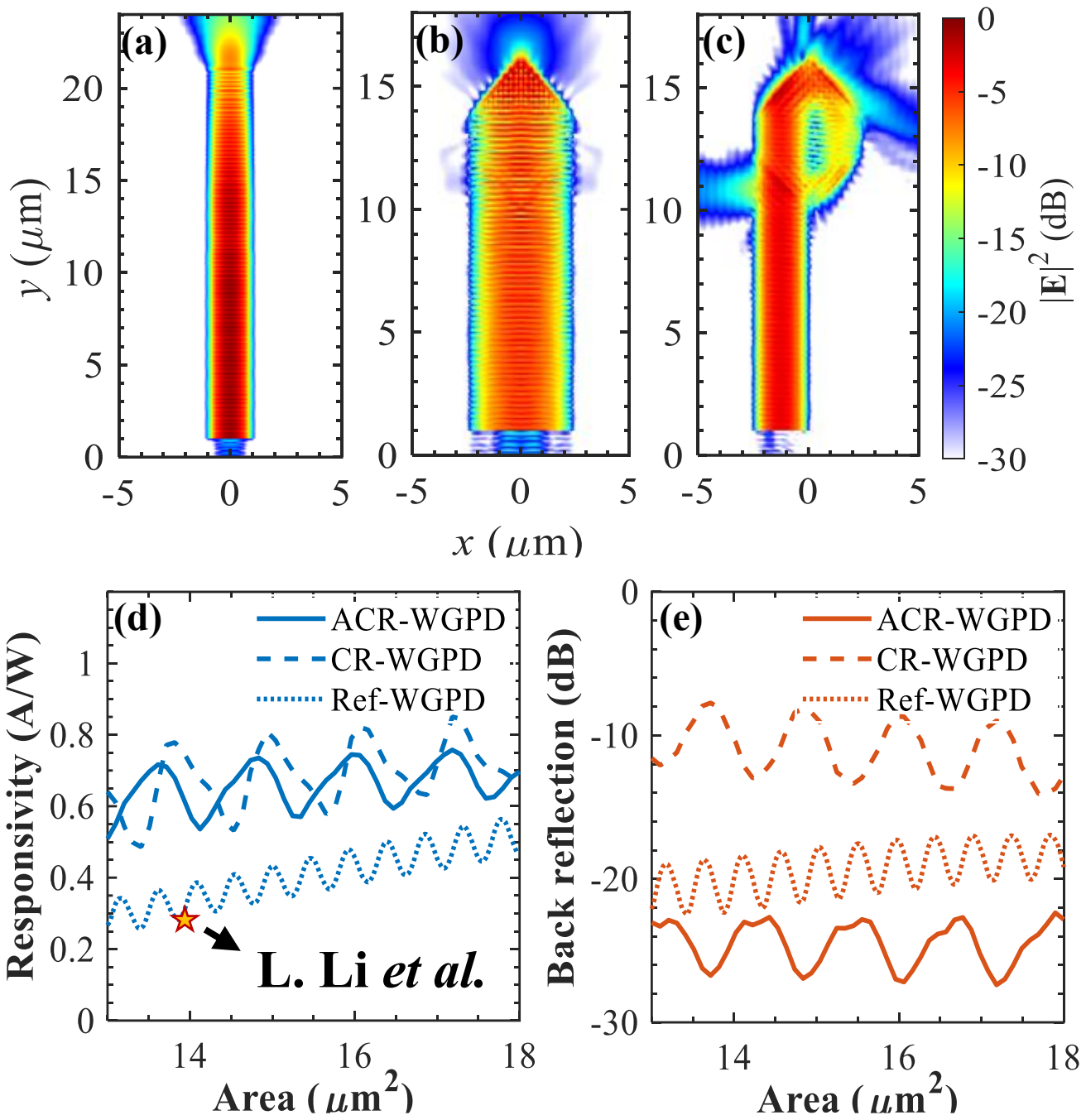

**FIG. 3**. Normalized optical-field intensity ($|\boldsymbol{E}|^2$) distributions in the waveguide layer for (a) the Ref-WGPD with awaveguide width of 2 μm, (b) the CR-WGPD with a waveguide width of 5 μm, and (c) the proposed ACR-WGPD with a waveguide entrance width of 2.5 μm. Calculated (d) responsivity and (e) optical back-reflection as functions of the active area over the range of 13~ 18 μm² for the three photodiode configurations.

To address insufficient absorption without increasing the active area, the symmetric CR-WGPD introduces a path-folding mechanism using a TIR corner structure. This geometry recycles unabsorbed light for multiple passes through the active region. Consequently, the responsivity increases from approximately 0.3 to 0.6 A/W, as shown in Fig. 3(d). However, this symmetric structure inherently acts as a retroreflector. It couples the recycled light directly back into the fundamental input mode. This drastically increases the optical back-reflection to range of -14 to -8 dB, as shown in Fig. 3(e). Such strong feedback can severely destabilize the external lasers required for co-packaged optics applications.

The proposed ACR-WGPD resolves this inherent trade-off by manipulating the reflected wave vector. It breaks geometric symmetry while retaining the path-folding mechanism. This maintains a high responsivity of approximately 0.6 A/W, which doubles that of the Ref-WGPD and matches the CR-WGPD, as shown in Fig. 3(d). Specifically, for a typical active area of 15 $\mu m^2$, the responsivity reaches 0.68 A/W. Coupled with the aforementioned 275 GHz bandwidth as shown in Fig.2, this yields a high bandwidth-efficiency product (BEP) of 150 GHz. Moreover, the asymmetric design redirects any remaining light laterally toward the sidewalls. This steers the residual light outside the angular acceptance range of the input waveguide. Consequently, it geometrically interrupts the longitudinal retroreflective path and suppresses mode coupling. As a result, the ACR-WGPD suppresses the optical back-reflection to between -28 and -23 dB, as shown in Fig. 3(e).

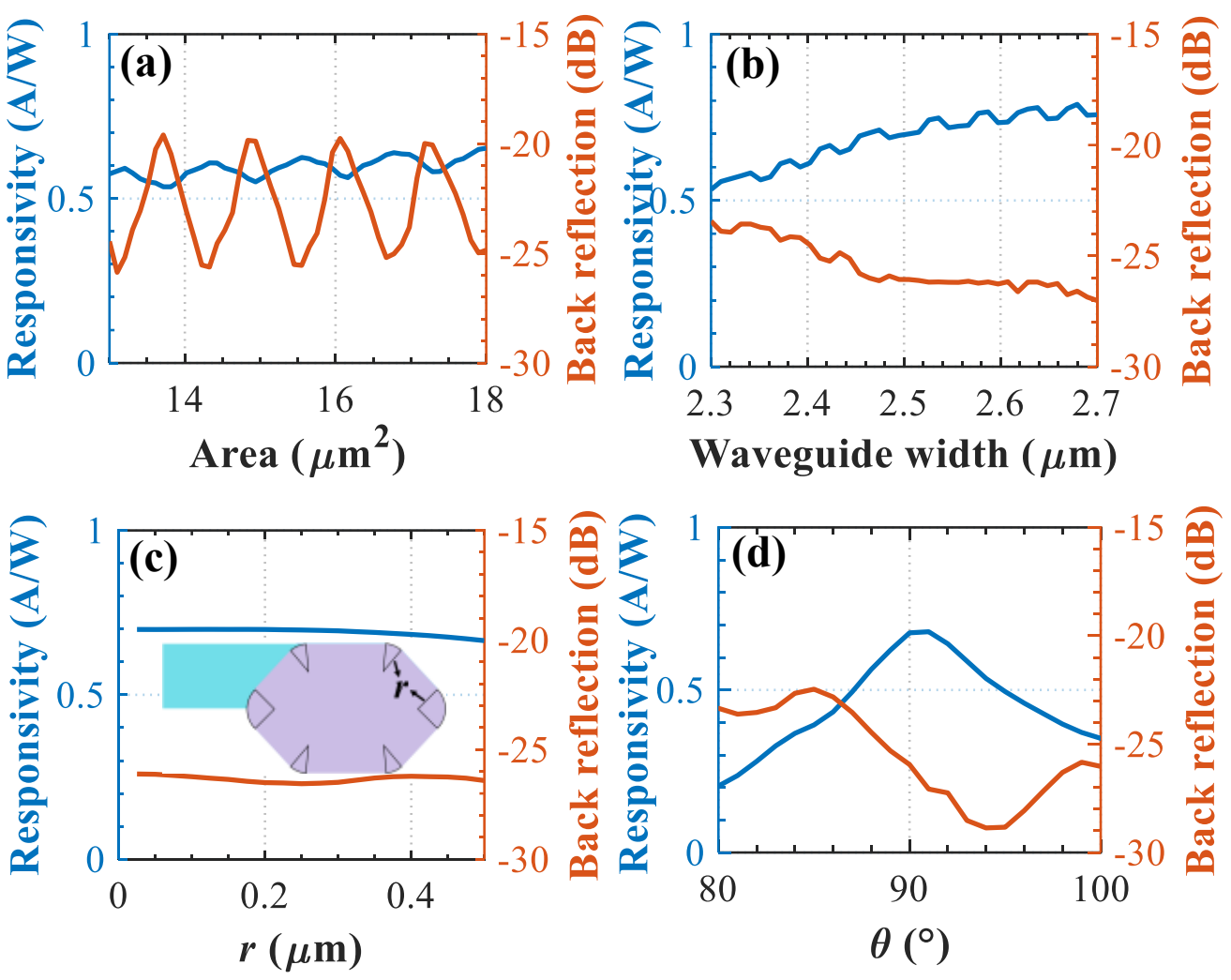

**FIG. 4.** Simulated responsivity (left vertical axis) and optical back-reflection (right vertical axis) as functions of four structural parameters: (a) the active area under TM mode excitation, (b) the waveguide width, (c) the corner curvature radius $\boldsymbol{r}$, and (d) the CR opening angle $\boldsymbol{\theta}$. The active area is fixed at 15 μm² for cases (c) and (d).

To assess the practical viability of the proposed ACR-WGPD, we evaluate its performance under different polarization states and its tolerance to fabrication variations. The responsivity under TM-mode excitation remains approximately 0.6 A/W, as shown in Fig. 4(a). This indicates that the TIR condition at the semiconductor-cladding interface is maintained for both polarizations. Note that the optical back-reflection for the TM mode ranges from -26 to -20 dB, slightly higher than that of the TE mode.

Standard semiconductor foundry processes, such as deep dry etching and photolithography, inherently introduce dimensional variations. The tolerance of the ACR-WGPD to these fabrication variations is evaluated in Figures 4(b) and 4(c). Specifically, the input waveguide width may deviate due to lateral over- or under-etching, as shown in Fig. 4(b). From an optical perspective, as waveguide width increases from 2.3 μm to 2.7 μm, the responsivity increases from approximately 0.55 A/W to 0.8 A/W, accompanied by a gradual decrease in the optical back-reflection. Physically, a wider waveguide increases the absorption volume, directly yielding a higher responsivity. Concurrently, the widened structure enables a more gradual transition of the effective refractive index. This smoother index profile minimizes optical discontinuities, thereby effectively suppressing parasitic reflection and scattering. Nevertheless, this optical benefit must be weighed against electrical constraints. Since the waveguide width dictates the active mesa size, widening it linearly increases junction capacitance and the RC time constant, thereby degrading the 3-dB RF bandwidth. Therefore, the nominal waveguide width requires careful optimization to balance high responsivity against RC-limited bandwidth requirements.

Furthermore, we evaluate the impact of lithography-induced corner rounding, as perfectly sharp vertices are practically unattainable, as shown in Fig. 4(c). As the radius of curvature ($r$) increases from 25 to 500 nm, both the responsivity and the optical back-reflection remain remarkably stable. This confirms that the ACR-WGPD performance is governed by the macroscopic asymmetry of the angled facets, rather than the microscopic sharpness of the vertices. Consequently, this geometric insensitivity ensures high fabrication tolerance, alleviating the strict requirement for ultra-high-resolution lithography.

Although highly tolerant to dimensional variations and corner rounding, the device is intrinsically sensitive to the corner-reflector opening angle ($\theta$), which dictates the folded optical trajectory. As shown in Fig. 4(d), A strong angular dependence is observed. The responsivity peaks for $\theta$ between 90° and 92°, indicating that a near-orthogonal corner provides the optimal lateral path for multipass absorption. Conversely, the optical back-reflection is minimized at slightly obtuse angles (93° to 95°), reaching a floor of approximately -29 dB. Physically, this obtuse angle compensates for the wavefront tilt induced by off-center edge propagation, maximizing modal misalignment to suppress optical feedback.

In conclusion, we have proposed and numerically evaluated an ACR-WGPD to effectively overcome the conventional bandwidth-responsivity limitation. Rather than compromising optical absorption in a miniaturized volume, the proposed design exploits a path-folding geometry to enable robust multipass light trapping. This geometric configuration allows the device to deliver a responsivity of 0.68 A/W and a 3-dB bandwidth of 275 GHz within a compact 15 μm² active area, these values correspond to a record-high BEP of 150 GHz for UTC-PDs, theoretically confirming that extreme device scaling can be achieved without efficiency degradation. Moreover, this asymmetric design geometrically suppresses back-reflection to -29 dB. It also maintains simulated robustness against typical fabrication tolerances, such as corner-rounding radii, corner-opening angles, and waveguide-width variations, while supporting both TE and TM modes. Consequently, leveraging its core advantages of a compact footprint, high bandwidth, and high responsivity, this architecture presents a highly practical and scalable solution for next-generation, high-density photonic integration.

This work was supported by the National Key Research and Development Program of China (Grant No. 2025YFE0203900), the Key R&D Program of Zhejiang (2025C01218), and the Open Research Fund of Beijing Key Laboratory of MEMS Technology and Device Reliability for the Industrial Internet (2025MKF01).

## AUTHOR DECLARATIONS

### Conflict of Interest

The authors have no conflicts to disclose.

## DATA AVAILABILITY

The data that support the findings of this study are available from the corresponding author upon reasonable request.